\documentclass[preprint]{raa}
\usepackage{amssymb}            
\usepackage{graphicx,times}
\usepackage{natbib}

\providecommand{\bm}[1]{\mbox{\boldmath$#1$\unboldmath}}

\def\etal{et al.}

\begin{document}

\title{A brief report on statistical study of net electric current in solar active regions
    with longitudinal fields of opposite polarity
 $^*$
\footnotetext{\small $*$ Supported by the National Natural Science
Foundation of China.} }

\volnopage{ {\bf ????} Vol.\ {\bf ?} No. {\bf XX}, 000--000}
   \setcounter{page}{1}

\author{Y. Gao\inst{1}}

\institute{Key Laboratory of Solar Activity, National Astronomical
Observatories, National Astronomical Observatories, Chinese Academy
of Sciences, Beijing 100012, China; {\it gy@bao.ac.cn}\\
\vs \no
   {\small Received [year] [month] [day]; accepted [year] [month] [day] }
}

\abstract{Dynamic processes occurring in solar active regions are
dominated by the solar magnetic field. As of now, observations using
a solar magnetograph have supplied us with the vector components of
a solar photospheric magnetic field. The two transverse components
of a photospheric magnetic field allow us to compute the amount of
electric current. We found that the electric current in areas with
positive (negative) polarity due to the longitudinal magnetic field
have both positive and negative signs in an active region, however,
the net current is found to be an order-of-magnitude less than the
mean absolute magnitude and has a preferred sign. In particular, we
have statistically found that there is a systematic net electric
current from areas with negative (positive) polarity to areas with
positive (negative) polarity in solar active regions in the northern
(southern) hemisphere, but during the solar minimum this tendency is
reversed over time at some latitudes. The result indicates that
there is weak net electric current in areas of solar active regions
with opposite polarity, thus providing further details about the
hemispheric helicity rule found in a series of previous studies.
}

\authorrunning{Y. Gao}            
\titlerunning{Net Current in Solar Active Regions}  
\maketitle


%
%
\section{Introduction}           
\label{sect:intro}

In the solar atmosphere, a chirality has been found in the
photospheric magnetic field of active regions that is usually
described by the parameter of current helicity or force-free field
factor. In the northern (southern) solar hemisphere, the helicity
mainly possess left (right) handedness, and is called the
hemispheric helicity sign rule (Seehafer 1990; Pevtsov et al. 1994;
Pevtsov et al. 1995; Abramenko et al. 1996; Bao \& Zhang 1998;
Hagino \& Sakurai 2004; Zhang et al. 2010). Futhermore, the
distribution of helicity signs was investigated between the
23$^{rd}$ and 24$^{th}$ solar activity cycles (e.g., Hao \& Zhang
2011). Generally, both left and right handed chirality coexist on a
particular pixel in an image of solar active region, or even a
sunspot, used in deriving a solar vector magnetogram. The physical
explanation of the hemispheric helicity sign rule is still a
complicated problem. Some typical case studies have shown that there
are systems with opposite net current in several active regions
(Wang et al. 1994; Leka et al. 1996; Wang \& Abramenko 1999;
Wheatland, 2000). However, it was also inferred that there was no
net current in sunspots (Venkatakrishnan \& Tiwari 2009). The
application of the parameter describing current helicity in active
regions has again come into question. In this paper, we try
separating the positive and negative flux areas according to the
longitudinal field of the solar active region and study the
distribution of current. In Section 2, we describe the procedure of
observation for this analysis, and present data reduction and result
in Section 3. We then summarize and discuss our results in Section
4.


\section{Observations}
\label{sect:Obs}

Vector magnetograms were observed using the Solar Magnetic Field
Telescope (SMFT), which has a tunable birefringent filter-type video
vector magnetograph (Ai 1987). For photospheric observations, the
passband of the filter is set at FeI $\lambda$5324 \AA{}. Normally
the longitudinal component of the field (Stokes V) is measured at
-75 m\AA{} and the transverse components (Stokes Q and U) are
measured at the line core. The equivalent width of the FeI
$\lambda$5324 \AA{} line is 0.344 \AA{} and the FWHM of the filter
passband is 0.15 \AA{}. Vector magnetograms are reconstructed by the
following relations:
\begin{equation}
B_{\parallel}=C_{\parallel}\frac{V}{I}
\end{equation}
\begin{equation}
B_{\perp}=C_{\perp}\left[\left(\frac{Q}{I}^{2}+\frac{U}{I}^{2}\right)
\right] ^{1/4}
\end{equation}
in which C$_{\parallel}$ and C$_{\perp}$ are the calibration
coefficients for the longitudinal and transverse fields (Su \& Zhang
2004). The azimuthal angle of the transverse field is
\begin{equation}
\phi=\frac{1}{2}\arctan \left(\frac{U}{Q} \right)
\end{equation}
Thus the $y$- and $x$- components of the vector magnetic field are
as follows:
\begin{equation}
B_{y}=B_{\perp}\sin\phi
\end{equation}
\begin{equation}
B_{x}=B_{\perp}\cos\phi
\end{equation}
The $180^{\circ}$ ambiguity in the direction of the transverse field
is resolved following Wang et al. (1994) by comparison with the
potential field. The active regions under investigation are all
located near the center of the disk. The latitude and longitude are
less than 40$^{\circ}$, so that the projection effects are small.
Therefore we may denote the line-of-sight and horizontal components
of magnetic field as $B_{z}$ and $B_{t}$, respectively. Usually, in
our computations, we select the pixels where the signal exceeds the
noise levels $|B_{z}|>20 G$ and $B_{t}>100 G$. The difference in the
method used for data reduction compared with previous research is
that we compute the average value of $J_{z}$ ($\langle J_{z}
\rangle$) on the pixels where ($B_{z} > 20 G$, $B_{t} > 100 G$) or
($B_{z} < -20 G$, $B_{t}
> 100 G$) respectively. The derived parameter is longitudinal
electric current, which is defined as $J_z=(1/\mu_{0}) (\partial
B_{y}/\partial x-\partial B_{x}/\partial y)$.

\section{Data reduction}
\label{sect:data}

\subsection{Case of NOAA 10484}

Figure 1 shows a simple relation between the electric current and
positions of the pixels. The top panel shows the electric current in
areas with both positive and negative longitudinal fields in the
active region NOAA 10484. The middle and bottom panels show the
electric current in areas with only positive or negative
longitudinal field respectively. Compared with the top panel, the
bottom two show a weak negative and positive deviation in electric
current respectively, in spite of all their standard deviations
being of the same order.


\begin{figure}
\centering
\includegraphics[angle=0,scale=.80]{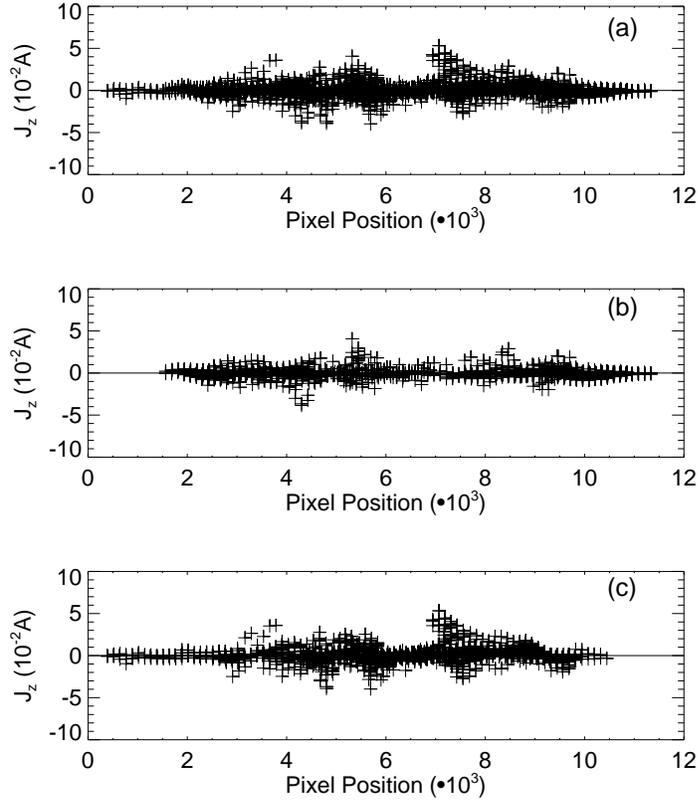}
\caption{Electric current as a function of pixel position. (a) shows
the electric current with both positive and negative longitudinal
fields in the active region NOAA 10484. The average and standard
deviation are 1.0 $\times 10^{-2}$ A and 6.6 $\times 10^{-1}$ A,
respectively. (b) shows the electric current in areas with only
positive longitudinal fields. The average and standard deviation are
-7.6 $\times 10^{-2}$ A and 5.0 $\times 10^{-1}$ A, respectively.
(c) shows the electric current in areas with only negative
longitudinal fields. The average and standard deviation are 7.4
$\times 10^{-2}$ A and 7.5 $\times 10^{-1}$ A, respectively.}
\end{figure}

Figure 2 shows the detailed information on the distribution of the
electric current of NOAA 10484. The vector magnetogram was observed
at 09:49:42 on 2003 Oct 22. There are three columns in Figure 2, the
first column shows the distribution of electric current in areas
with both positive and negative longitudinal fields. The color
patterns in row 2 - 4 show the $J_{z}$ in intervals with an
order-of-magnitude $<$ 10$^{0}$ to 10$^{-4}$ respectively. The
number of pixels where $|J_{z}| > 10^{0}$ A accounts for 2.5\% of
the total. The number of pixels where $|J_{z}| < 10^{-3}$ A accounts
for 57\% of the total, but the $\langle |J_{z}| \rangle$ is only on
the order of $10^{-5}$, which is not shown in this figure. The main
contribution of $\langle |J_{z}| \rangle$ has an order-of-magnitude
of $10^{-3}-10^{0}$ A, as shown in the Row 2-4 of the first column.


\begin{figure}
\centering
\includegraphics[angle=0,scale=.70]{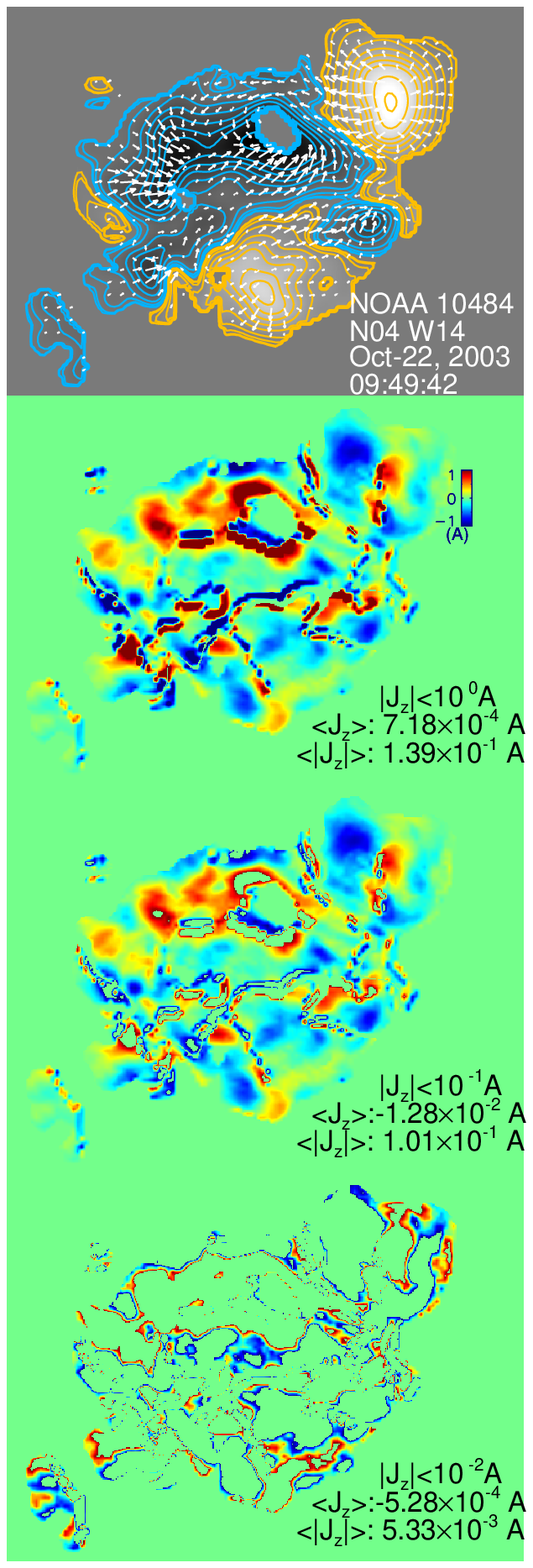}
\includegraphics[angle=0,scale=.70]{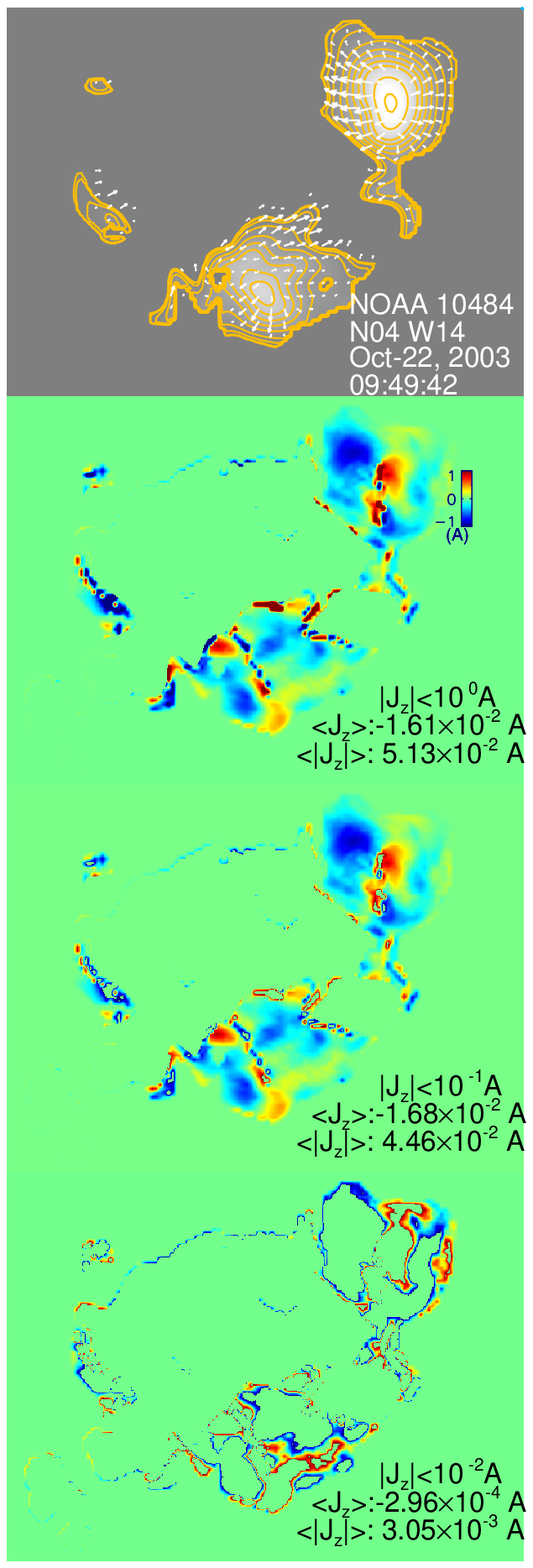}
\includegraphics[angle=0,scale=.70]{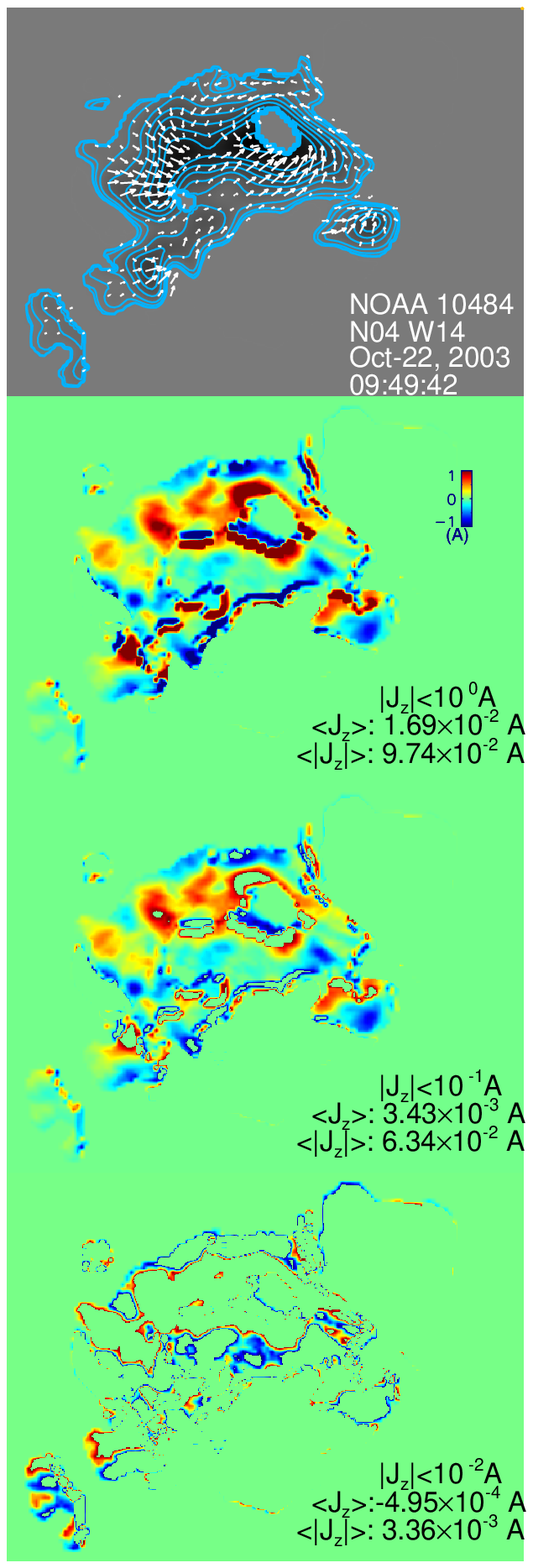}
\caption{The top panels in each column show the original vector
magnetogram and the gray-scale images show the longitudinal magnetic
field of NOAA 10484. The orange (cyan) contours show the positive
(negative) longitudinal magnetic field. The contour levels are
[¡À30, 160, 480, 800, 1280, 1600G]. The first column shows the
distribution of electric current in areas with both positive and
negative longitudinal fields. The second column shows the
distribution in the area with only a positive longitudinal field.
The third column shows the distribution in the area with only a
negative longitudinal field. The images from the second to fourth
rows show the electric current. The corresponding scale of magnitude
is shown in the bottom right corner of each plot.}
\end{figure}

The second column shows the distribution in areas with only a
positive longitudinal field. It is found that a mean current -1.61
$\times 10^{-2}$, -1.68 $\times 10^{-2}$ and  -2.96 $\times 10^{-4}
$ A with a magnitude smaller than the mean absolute magnitude of
5.13 $\times 10^{-2}$, 4.46 $\times 10^{-2}$ and  3.05 $\times
10^{-3} $ A corresponding to the orders of $10^{0}$, $10^{-1}$ and
$10^{-2}$ A, respectively.

The third column shows the distribution in areas with only a
negative longitudinal field. It is found that there is a mean
current of 1.69 $\times 10^{-2}$, 3.43 $\times 10^{-3}$ and -4.95
$\times 10^{-4}$ A with a magnitude smaller than the mean absolute
magnitude of 9.74 $\times 10^{-2}$, 6.34 $\times 10^{-2}$ and 3.36
$\times 10^{-3}$ A corresponding to the order of $10^0$, $10^{-1}$
and $10^{-2}$, respectively.

\subsection{Statistic Study of Large Sample}

We analyze a database of 6629 vector magnetograms observed at
Huairou Solar Observing Station from 1988 to 2005. For more detailed
information, the reader is referred to Zhang et al. (2010). The
effective field of view is 5.23$^{\prime}$ $\times$ 3.63$^{\prime}$
(before 2001 Aug 25), 4.06$^{\prime}$ $\times$ 2.77$^{\prime}$ (2001
Aug 25 to 2001 Oct 13 and 2001 Oct-14 to Nov-30) and 3.75$^{\prime}$
$\times$ 2.81$^{\prime}$ (after 2001 Dec 1).  After obtaining
$\langle J_{z} \rangle$ for each magnetogram, we compute the 2-year
running average value in a 7$^{\circ}$ range of latitudes. Then we
derive butterfly diagrams, as shown in the figure 3.


\begin{figure}
\centering
\includegraphics[angle=90,scale=.60]{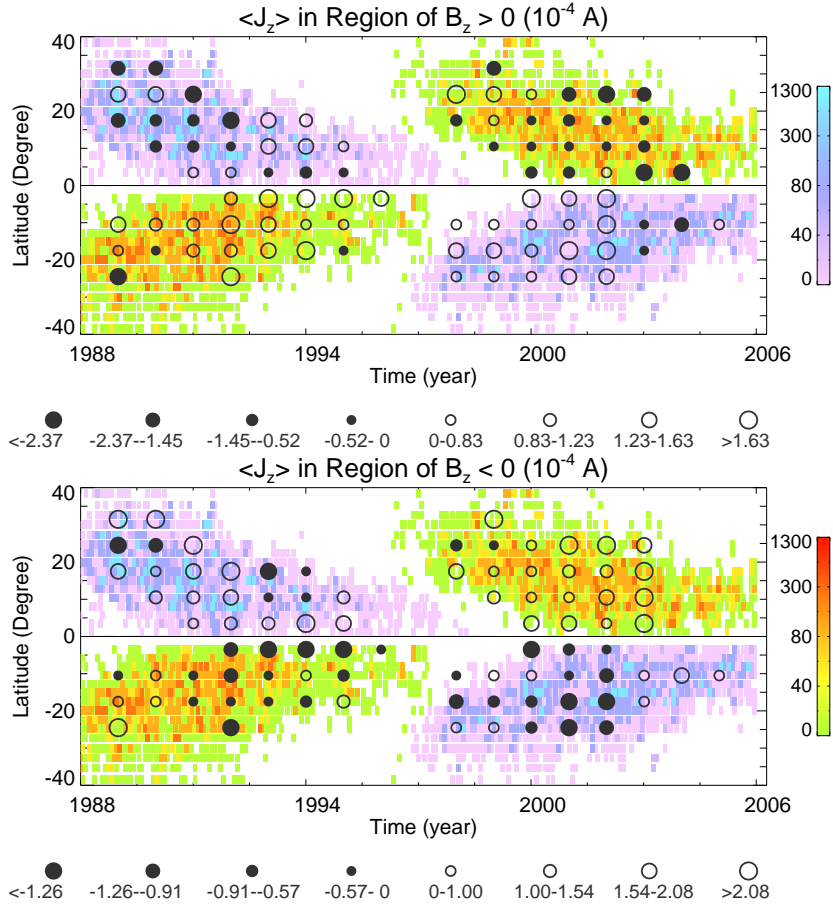}
\caption{The solid circles (\textit{filled circles}) show the
positive (negative) sign of the mean electric current. The sizes of
the circles are proportional to the magnitude of the displayed
quantity. The underlying colored ``butterfly diagram¡± shows how
sunspot density varies with latitude over the solar cycle. The upper
(lower) vertical color bar on the right represents the leading
sunspot having negative (positive) polarity.}
\end{figure}

Figure 3 shows that the $\langle J_{z} \rangle$ is negative
(positive) in areas with positive (negative) magnetic polarity in
northern hemisphere or positive (negative) in areas with positive
(negative) magnetic polarity in the southern hemisphere in most
bins. The colored background is the butterfly diagram of sunspot
areas. The vertical colored bar beside the upper (lower) panel
indicates that the leading polarity is mainly negative (positive)
and the following polarity is mainly positive (negative). So the
whole butterfly diagram statistically indicates that the direction
of the weak net current points from sunspots with negative
(positive) polarity to ones with positive (negative) polarity in the
northern (southern) hemisphere. Additionally, there are obvious
exceptions at the minimum of the solar activity cycle. This implies
that a statistical reversal in the sign of helicity at the minimum
of the solar activity cycle is an intrinsic property of evolution of
the solar magnetic field. It should be pointed out that exceptions
are found not only in the minimum but also in the whole solar
activity cycle. The percentage of exceptions is relatively higher at
the minimum than in other phases of the solar cycle.

Figure 4 shows two exceptions. The upper panels show a vector
magnetogram of NOAA 9690 (left) where the distribution of electric
current by order-of-magnitude is less than $10^{0}$ A (right). The
$\langle J_{z} \rangle$ is $-1.22 \times 10^{-2} A$ compared to the
$\langle |J_{z}| \rangle$ of $1.08 \times 10^{-1} A$ (11.3\%). The
bottom panels show the vector magnetogram of NOAA 8675 (Left) and
the distribution of electric current by order-of-magnitude is less
than $10^{0} $A (right). The $\langle J_{z} \rangle$ is $-2.06
\times 10^{-3} A$ compared to the $\langle |J_{z}| \rangle$ of $1.86
\times 10^{-2} A$ (11.1\%).


\begin{figure}
\centering
\includegraphics[angle=0,scale=1.0]{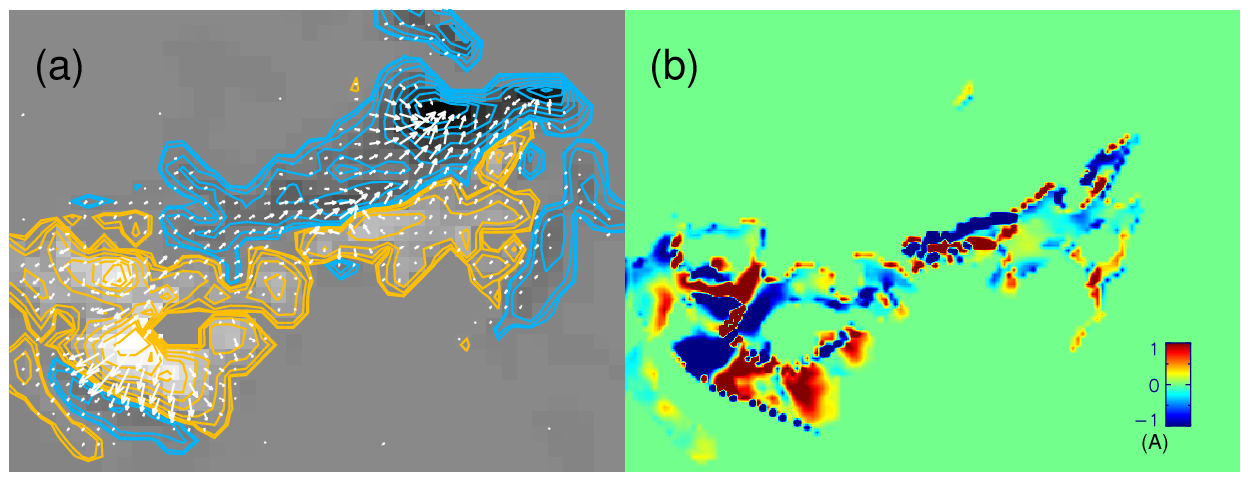}
\includegraphics[angle=0,scale=1.0]{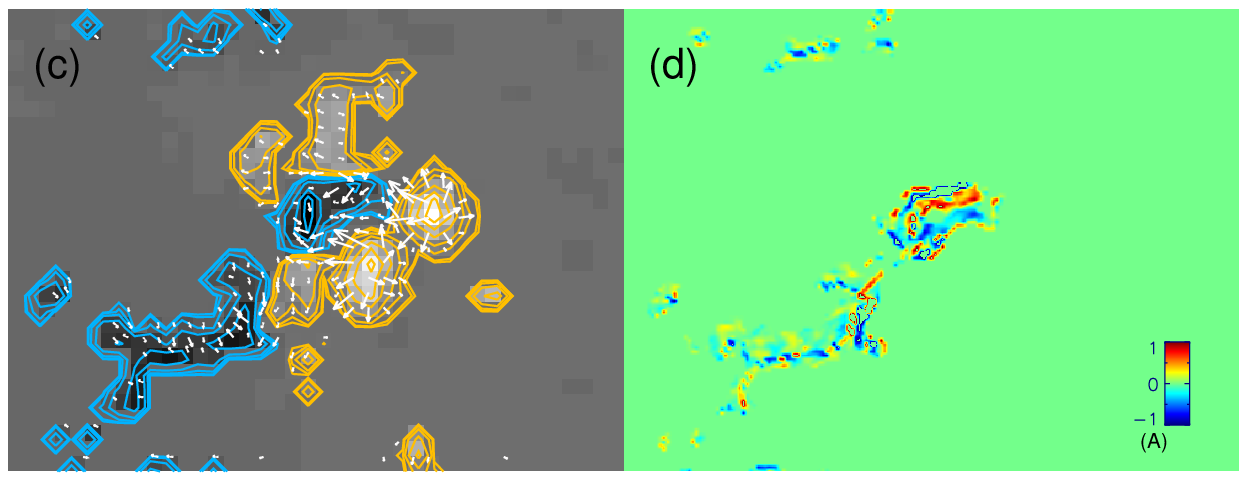}
\caption{(a) The upper panels show a vector magnetogram of NOAA 9690
which was located at S17E31 and observed at 11:38:58 on 2001 Nov 08
and (b) areas where the distribution of electric current is in an
area with positive polarity. (c) The bottom panels show a vector
magnetogram of NOAA 8657 which was located at N19W12 and observed at
02:50:01 on 1999 Aug 31 and (d) areas where the distribution of
electric current is in an area with negative polarity.}
\end{figure}

\section{CONCLUSIONS AND DISCUSSION}
\label{sect:conclusion}

Through analyzing magnetograms, we have statistically found that
there is a systematic net electric current from areas with negative
(positive) polarity to areas with positive (negative) polarity in
solar active regions in the northern (southern) hemisphere. The tiny
net electric current ($\langle J_{z} \rangle$) is an
order-of-magnitude smaller than the mean absolute magnitude of
electric current $\langle |J_{z}| \rangle$ in the whole active
region. This current reveals more details about the hemispheric
helicity sign rule found in a series of previous studies. For
example, the net current is statistically negative in areas with
positive (negative) polarity for solar active regions in the
northern (southern) hemisphere. As a result, this will cause an
excess negative (positive) current helicity in the northern
(southern) hemisphere, which is consistent with the sign preference
associated with the hemispheric helicity sign rule.

At the minimum in the solar activity cycle, some exceptions of net
electric current have been found, which have an opposite direction
with respect to the normal ones. This gives self-consistent evidence
that the reversal sign of helicity based on observations from the
Huairou Solar Observing Station reflects the intrinsic property of
twist in the solar magnetic field. Furthermore, we compute the
percent of active regions conforming or not conforming to the
hemispheric helicity sign rule.

As given in Table 1, bold text shows the cases that follow the
overall pattern. Besides a weak preference in sign, there are some
unusual features. For the 22$^{nd}$ solar cycle, the percentages are
67.6\% (61.9\%) when $B_z$ $>$ 0 ($B_z$ $<$ 0) in the southern
(northern) hemisphere. The percentages are higher than 51.6\%
(46.3\%) when $B_z$ $<$ 0 ($B_z$ $>$ 0) in the southern (northern)
hemisphere. For the 23$^{rd}$ solar cycle, the percentages are
62.7\% (67.8\%) when $B_z$ $>$ 0 ($B_z$ $<$ 0) in the southern
(northern) hemisphere. The percentages are still higher than 51.1\%
(53.1\%) when $B_z$ $<$ 0 ($B_z$ $>$ 0) in the southern (northern)
hemisphere. This imbalance of net current in different hemispheres
when $B_z$ $<$ 0 or $B_z$ $>$ 0 may arise from the effect of Faraday
rotation, which can cause an excess in false net positive current in
an active region (Hagino and Sakurai, 2004; Gao et al., 2008).
Nevertheless, the areas with negative current in the southern
(northern) hemisphere of $B_z$ $>$ 0 ($B_z$ $<$ 0) show convincing
features that support the reversal in the sign helicity being
independent of the effect of Faraday rotation. Although the
hemispheric helicity sign rule holds over the whole 23$^{rd}$ solar
activity cycle, the apparent areas of reversal sign are seen in the
southern hemisphere at the end of the cycle. This is in agreement
with another statistical analysis done by Hao and Zhang (2011) with
independent data observed by Hinode in the descending phase of
23$^{rd}$ solar activity cycle.

\begin{table*}
  \begin{tabular}{lllllll}\hline \\

Hemisphere & ($\langle J_{z} \rangle$ $>$ 0, $B_z$ $>$ 0)  &
($\langle J_{z} \rangle$ $<$ 0, $B_z$ $>$ 0) & ($\langle J_{z}
\rangle$ $>$ 0, $B_z$ $<$ 0)  & ($\langle J_{z} \rangle$ $<$ 0,
$B_z$ $<$ 0)  \\ \hline\\
&& \multicolumn{1}{c}{$22^{nd}$ cycle} &&&\\ \hline \\
\multicolumn{1}{c}{North}  & \multicolumn{1}{c}{53.6\%}  &    \multicolumn{1}{c}{\textbf{46.3\%}}      &  \multicolumn{1}{c}{\textbf{61.9\%}}   &\multicolumn{1}{c}{38.1\%}          \\
 \multicolumn{1}{c}{South}  &   \multicolumn{1}{c}{\textbf{67.6\%}}  &   \multicolumn{1}{c}{32.4\%}      &  \multicolumn{1}{c}{48.4\%}   &\multicolumn{1}{c}{\textbf{51.6\%} }         \\ \hline\\
&& \multicolumn{1}{c}{$23^{rd}$ cycle} &&&\\ \hline \\
 \multicolumn{1}{c}{North}  &   \multicolumn{1}{c}{48.9\%}  &   \multicolumn{1}{c}{ \textbf{51.1\%}}      &  \multicolumn{1}{c}{\textbf{67.8\%}}  &   \multicolumn{1}{c}{ 32.3\%}    \\
 \multicolumn{1}{c}{South}  &  \multicolumn{1}{c}{\textbf{62.7\%}}   & \multicolumn{1}{c}{37.3\%}    &  \multicolumn{1}{c}{46.9\%}   & \multicolumn{1}{c}{\textbf{53.1\%} }       \\ \hline
\end{tabular}
\caption{The percentages of active regions following (bold text) or
not following to the pattern of hemispheric helicity sign rule.}
\end{table*}

\normalem
\begin{acknowledgements}

I am grateful to the referee for helpful comments and suggestions.
The work is supported by the National Natural Science Foundation of
China (Grant Nos. 11273034, 11178005, 41174153, 11173033, 11103037),
the National Basic Research Program of China (973 program,
2011CB811401), and Chinese Academy of Sciences under Grant
KJCX2-EW-T07.
\end{acknowledgements}

\label{lastpage}

\clearpage


\begin{thebibliography}{99}


\bibitem[Abramenko, Wang \& Yurchishin (1996)]{Abramenko96}
{Abramenko, V. I., Wang, T. J. \& Yurchishin, V. B.} 1996,
\textit{Solar Phys.}, 168, 75

\bibitem[Ai (1987)]{Ai87} {Ai, G.} 1987, Publ. Beijing Astron. Obs., No 9, 27-36

\bibitem[Bao \& Zhang (1998)]{BaoZ98} {Bao, S. D. \& Zhang, H. Q.} 1998,
\textit{ApJ}, 496, L43.

\bibitem[Gao, Su, Xu \& Zhang (2008)]{GSXZ08} {Gao, Y., Su, J.; Xu, H. \& Zhang, H.} 2008,
\textit{MNRAS.}, 386., 1959G.

\bibitem[Hagino \& Sakurai (2004)]{HS04} {Hagino, M. \& Sakurai, T.} 2004,
\textit{PASJ}, 56, 831.

\bibitem[Hao \& Zhang (2011)]{HZ11} {Hao, J. \& Zhang, M.} 2011,
\textit{ApJ}, 733, L27.

\bibitem[Leka \etal\ (1996)]{Leka96} {Leka, K. D., Canfield, R. C., McClymont, A. N. \& van Driel-Gesztelyi, L.} 1996,
\textit{ApJ}, 462, 547.

\bibitem[Pevtsov \etal\ (1994)]{Pevtsov94} {Pevtsov, A. A., Canfield, R. C. \& Metcalf, T. R.} 1994,
\textit{ApJ}, 425, L117.

\bibitem[Pevtsov \etal\ (1995)]{Pevtsov95} {Pevtsov, A. A., Canfield, R. C. \& Metcalf, T. R.} 1995,
\textit{ApJ}, 440, L109.

\bibitem[Seehafer (1990)]{Seehafer90} {Seehafer, N.} 1990,
\textit{Solar Phys.}, 125, 219.

\bibitem[Su \& Zhang (2004)]{SZ04} {Su, J. T. \& Zhang, ~H.} 2004,
\textit{Chinese Astronomy and Astrophysics}, \textbf{4}, 365;

\bibitem[Venkatakrishnan \& Tiwari (2009)]{Venkatakrishnan09} {Venkatakrishnan, P. \& Tiwari,
S. K.} 2009, \textit{ApJ}, 706, L114.

\bibitem[Wheatland, (2000)]{Wheatland00} {Wheatland, M. S.} 2000,
\textit{ApJ}, 532, 616.

\bibitem[Wang, Xu \& Zhang (1994)]{WangT94} {Wang, T. J., Xu, A. A. \& Zhang H. Q.} 1994,
\textit{Solar Phys.}, 155, 99.

\bibitem[Wang \& Abramenko (1999)]{WangT99} {Wang, T. J. \& Abramenko, V. A.} 1999,
\textit{ESASP.}, 448, 671.

\bibitem[Zhang \etal\ (2010)]{Zhang10} {Zhang, H. Q., Sakurai, T., Pevtsov, A., Gao, Y., Xu, H. Q., Sokoloff, D. D. \& Kuzanyan, K.} 2010,
\textit{MNRAS}, 402, L30.


\end{thebibliography}
\end{document}